\documentclass{aa}
\usepackage{graphicx}
\usepackage{txfonts}

\begin{document}
  \title{ Extrasolar planets and brown dwarfs around A-F type stars
  \thanks{ Based on observations made with the {\small HARPS}
  spectrograph at La Silla Observatory (ESO, Chile), and with the
  {\small CORALIE} spectrograph at La Silla Observatory (Swiss
  telescope).}}
  
  \subtitle{III. $\beta$~Pictoris : looking for planets, finding pulsations.}
  
  \author{
    F. Galland \inst{1,2}
    \and
    A.-M. Lagrange \inst{1}
    \and
    S. Udry \inst{2}
    \and
    A. Chelli \inst{1}
    \and
    F. Pepe \inst{2}
    \and
    J.-L. Beuzit \inst{1}
    \and
    M. Mayor \inst{2}
  }
  
  \offprints{
    F. Galland,\\
    \email{Franck.Galland@obs.ujf-grenoble.fr}
  }

  \institute{
    Laboratoire d'Astrophysique de l'Observatoire de Grenoble,
    Universit\'e Joseph Fourier, BP 53, 38041 Grenoble, France
    \and
    Observatoire de Gen\`eve, 51 Ch. des Maillettes, 1290 Sauverny, Switzerland
  }
  
  \date{Received 20 August 2005 / Accepted 23 September 2005}
  
  \abstract{
    In the frame of the search for extrasolar planets and brown dwarfs
    around early-type stars, we present the results obtained on
    $\beta$~Pictoris, which is surrounded by a circumstellar
    disk that is warped by the presence of a planet.
    We used 97 spectra acquired with {\small CORALIE} and
    230 spectra acquired with {\small HARPS} to
    characterize the radial velocity behavior of $\beta$~Pictoris and to
    infer constraints on the presence of a planet close to this
    star. With these data, we were able to exclude the presence of an
    inner giant planet (2~M$_{Jup}$ at a distance to the star of 0.05~AU,
    9~M$_{Jup}$ at 1~AU).
    We also discuss the origin of the observed radial velocity
    variations in terms of $\delta$~Scuti type pulsations.

  }
  
  \maketitle
  
  \section{Introduction}

  $\beta$~Pictoris (A5V, 19 pcs, \cite{crifo97}, $\approx$ 20 Myrs,
  \cite{Barrado99}) has been the subject of intensive
  investigations since the first discovery of an extended ($\geq$ 100
  AUs) circumstellar disk (\cite{ST84}) and since evidence that the
  lifetime of the grains in the disk was significantly shorter than
  the star age. It was then deduced that some grains were permanently
  formed through collisions among larger, possibly kilometer sized
  bodies, or perhaps by slow evaporation - at least partly (see
  \cite{lecavelier96}). The $\beta$~Pictoris disk was then considered
  as the first example of a resolved outer planetary system in a still
  unkown stage of evolution. Given the star age, it was possible that
  planets could be already formed or still under formation.
  Observation of a warp in the inner part of the disk was attributed
  to gravitational perturbation of the disk by a giant planet
  whose location could be constrained (\cite{mouillet97},
  \cite{augereau01} and references therein). Besides, episodes of
  strong and rapid infalls of ionized gas were detected and attributed
  to the evaporation of cometary objects grazing the star
  (\cite{Ferlet87}, \cite{lagrange88}, \cite{beust90}). Again, one or
  two giant planets within a few AUs were found to be necessary to
  trigger this infall of cometary bodies towards the star
  (\cite{beust96}, 2000). Finally, photometric variations were also
  detected once and possibly (but no exclusively) attributed to the
  presence of a planet passing the line of sight
  (\cite{lecavelier95}, 1997). For a review of these possible
  pieces of evidence, see e.g. \cite{Majar98} or \cite{LBA00} (2000).
  
  Direct detection of planets within a few AUs of a star aged 20
  Myrs or more is beyond the capability of current
  instrumentations. On the other hand, indirect detection through,
  e.g., radial velocity searches have been restricted to solar type
  stars until recently. Given the interest in understanding the planet
  formation process over a wide range of stellar characteristics and
  especially for massive stars, we set up a radial velocity survey
  dedicated to the search for planets around A-F type stars, using a
  dedicated analysis package that allows detection of companions
  down to planetary masses around such objects (\cite{Galland05a},
  Paper\,I). Here, we present the results of a radial velocity survey
  of $\beta$~Pictoris with {\small CORALIE} and {\small HARPS}
  performed over a period of several months. The data and the radial
  velocities obtained are presented in Sect.~2: the radial
  velocities are significantly variable. We show in Sect.~3 that
  these variations cannot be attributed to the presence of a
  planet. Sect.~4 explores other possible origins: stellar
  or cometary related. The origin of the variations finally involves
  pulsations of $\delta$~Scuti type. In Sect.~5, constraints are put
  on the remaining possible characteristics for a planet around
  $\beta$~Pictoris, taking the new constraints presented
  in this paper into account.
  
  \begin{figure*}[t!]
    \centering
    \includegraphics[width=0.75\columnwidth]{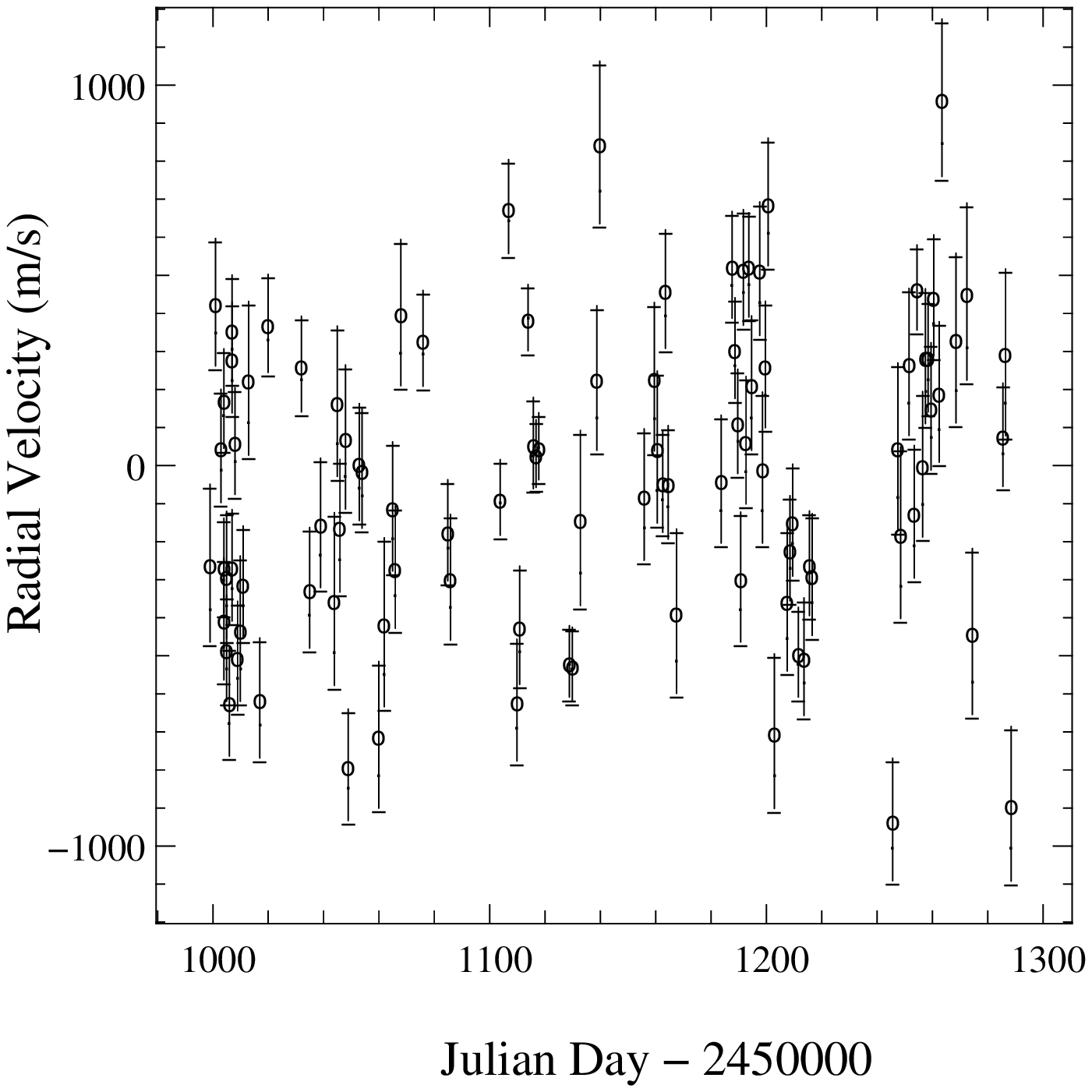}
    \includegraphics[width=0.75\columnwidth]{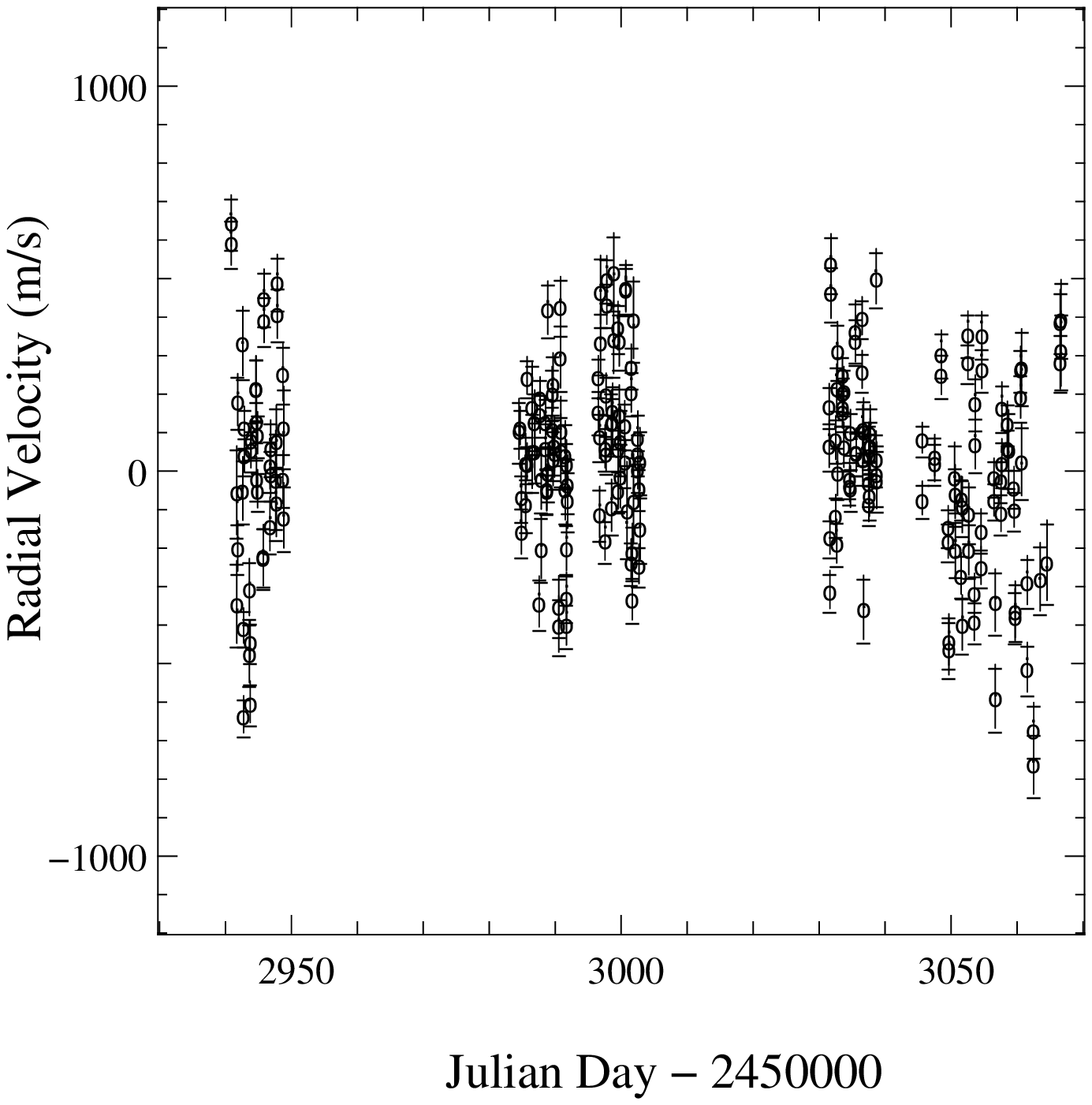}
    \includegraphics[width=0.75\columnwidth]{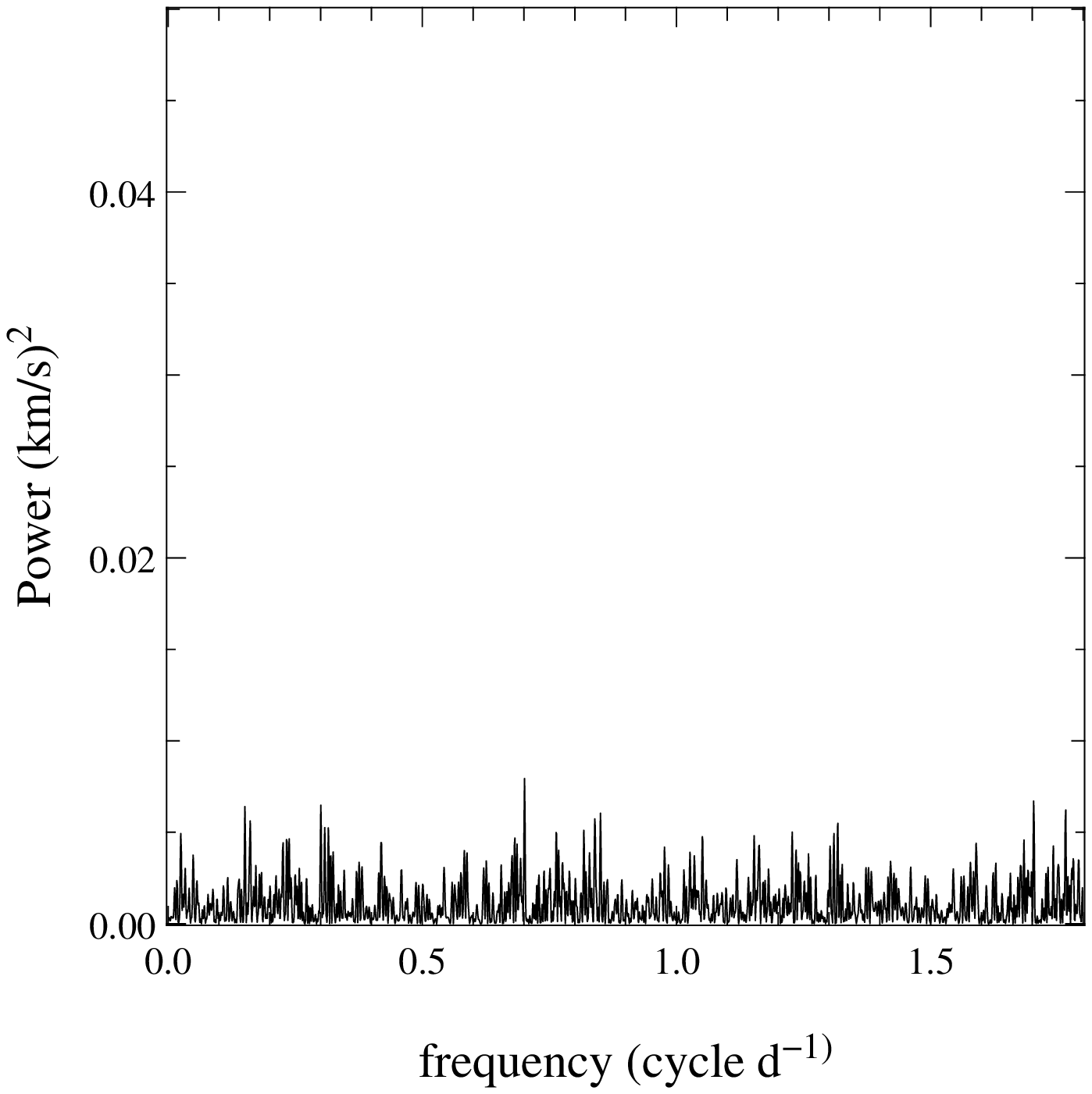}
    \includegraphics[width=0.75\columnwidth]{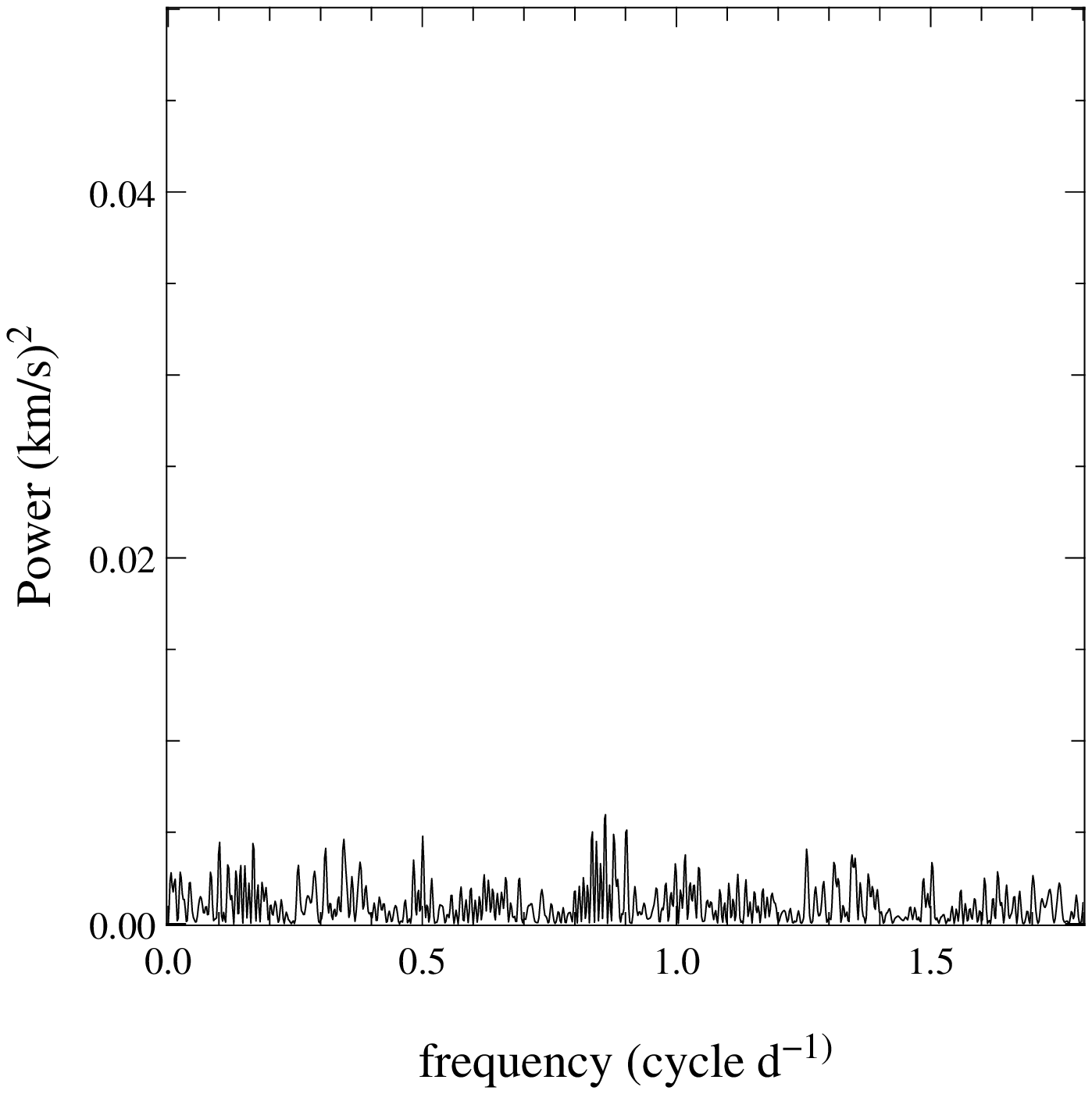}
    \caption{Radial velocity measurements (top) and related
    periodograms (bottom), obtained on $\beta$~Pictoris using the
    {\small CORALIE} (left) and {\small HARPS} (right)
    spectrograph. The scale of the periodograms is the same as in
    Fig.\,\ref{Har_dvft-perio_HF}; we used the CLEAN algorithm (see
    text) for these periodograms.}
    \label{Har_dvft-perio}
  \end{figure*}

  \section{Radial velocity measurements}
  \subsection{CORALIE}
  We acquired 120 spectra of $\beta$~Pictoris with the {\small
  CORALIE} spectrograph
  attached to the 1.2 m Swiss telescope at La Silla between
  July 1998 and April 1999 with a resolution R~$\approx$~50\,000. Of these,
  23 spectra with significantly lower S/N were rejected. We then
  considered the 97 spectra left with a mean S/N of $120$. 
  Each spectrum is composed of 68 spectral orders covering the wavelength range
  3900\AA \,to 6800\AA.
 
  For each spectrum, we selected 32 spectral orders containing deep lines,
  yet avoiding the strong \ion{Ca}{ii} and H lines, as well as
  the orders contaminated by telluric absorption lines. The radial velocities
  were measured using the method described in \cite{Chelli00} and
  in Paper\,I. They are displayed on Fig.\,\ref{Har_dvft-perio} (top,
  left). The individual uncertainty is 163\,m\,s$^{\rm -1}$ on average.

  \subsection{HARPS}
  We acquired 258 spectra of $\beta$~Pictoris with the {\small HARPS}
  spectrograph during period P73, between November 2003 and March 2004, with a
  resolution R~$\approx$~100\,000. The 29 spectra
  with lower S/N have been left over. We then considered the
  229 spectra left, with a mean S/N of $330$ (exposure time of around
  1 minute). Each spectrum is formed by 72 spectral orders covering
  the spectral window [3800\AA, 6900\AA].

  We performed the same treatement as for {\small CORALIE} and
  obtained the radial velocities displayed in
  Fig.\,\ref{Har_dvft-perio} (top, right). The individual
  uncertainty is 65\,m\,s$^{\rm -1}$ on average, which is consistent with
  the value of 60\,m\,s$^{\rm -1}$ obtained from simulations in
  Paper\,I by applying the relation between the radial velocity
  uncertainties and $v\sin{i}$ to $\beta$~Pictoris, with $\mathrm{S/N}$
  values equal to 330.

  \section{No inner giant planet}

  In the case of {\small CORALIE}, the dispersion of the measured radial
  velocities is 390\,m\,s$^{\rm-1}$ rms, i.e. a factor 2.4 higher than
  uncertainties. Radial velocities are thus significantly variable,
  even if the dispersion is still close to the uncertainties.

  The periodogram of the {\small CORALIE} radial velocities does not
  show any clear peak in a period range of 1-300 days
  (Fig.\,\ref{Har_dvft-perio}, bottom, left).  The observed radial
  velocity variations are thus not due to the presence of a planet.
  Note that we used the CLEAN algorithm (\cite{Roberts87}) in order to
  remove the aliases associated with temporal sampling of the data.
  This algorithm deconvolves the window function iteratively from the
  initial ``dirty'' spectrum to produce the resulting cleaned
  periodogram; the power obtained at a given frequency is the square
  of the radial velocity semi-amplitude of the corresponding potential
  radial velocity periodic variations.  Assuming a circular orbit,
  the data exclude the presence of a planet with a period lying
  typically between 1 and 600 days (hence a separation between 0.03
  and 1.8 AU) and with an induced radial velocity semi-amplitude larger
  than~$\approx$~400\,m\,s$^{\rm-1}$ (Fig.\,\ref{Diag-exclus}).

  In the case of {\small HARPS}, the dispersion of the measured radial
  velocities is 252\,m\,s$^{\rm-1}$ rms, i.e. a factor 3.9 higher than
  uncertainties; so, the radial velocities are really significantly
  variable. Assuming the same level of radial velocity variations at
  the time of {\small CORALIE} and {\small HARPS} observations, this
  higher factor with {\small HARPS} could be explained by its greater
  stability.

  The periodogram of the {\small HARPS} radial velocities does not
  show any clear peak in a period range of 1-180 days
  (Fig.\,\ref{Har_dvft-perio}, bottom, right). The observed radial
  velocity variations are thus not due to the presence of a
  planet. Assuming a circular orbit, we can exclude the presence of a
  planet with a period lying typically between 1 and 350 days (hence a
  separation between 0.03 and 1.2 AU) and with an induced radial velocity
  semi-amplitude larger than~$\approx$~250\,m\,s$^{\rm-1}$
  (Fig.\,\ref{Diag-exclus}).

  \begin{figure}[t!]
    \centering
    \includegraphics[width=0.8\columnwidth]{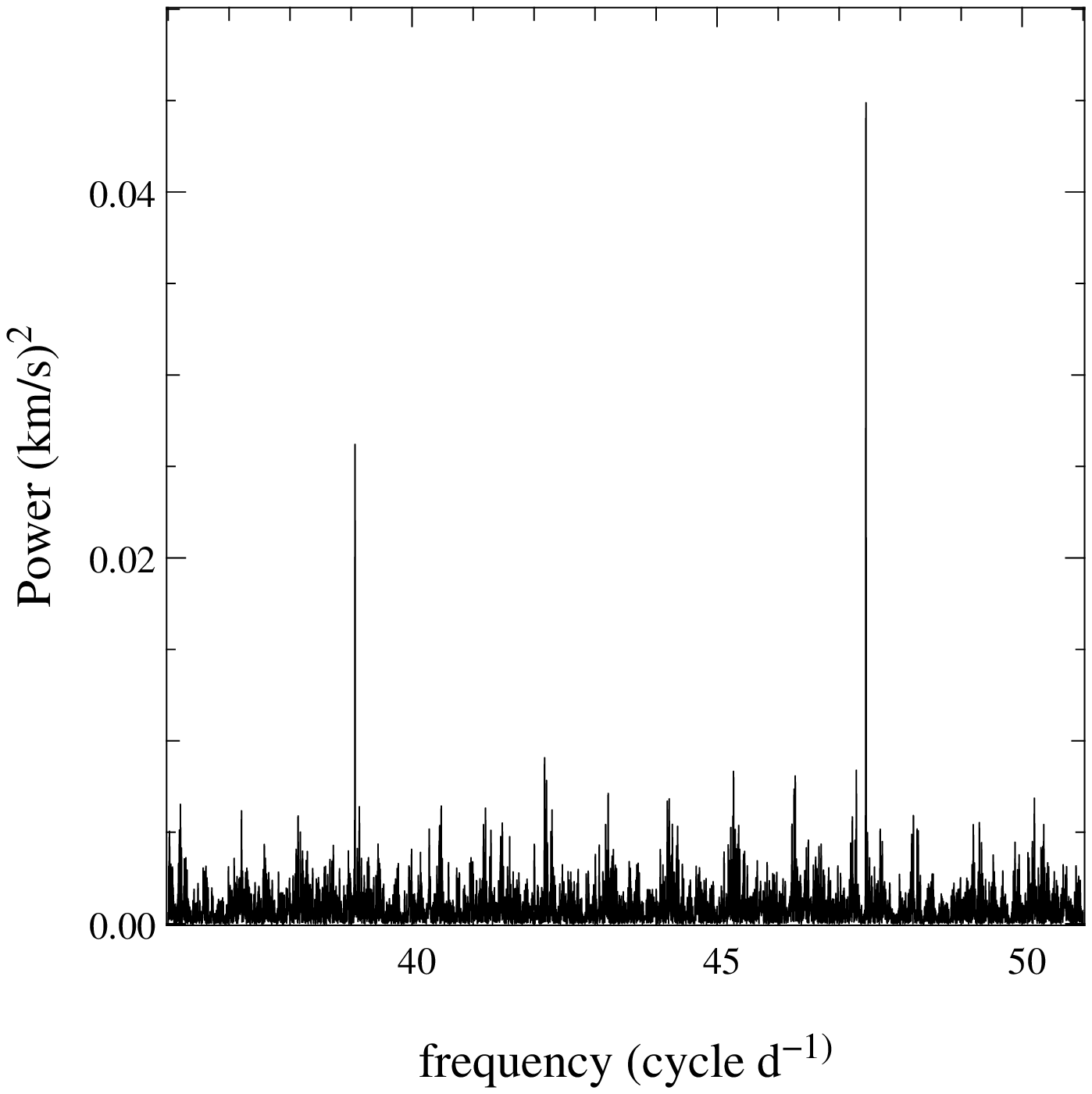}
    \includegraphics[width=0.8\columnwidth]{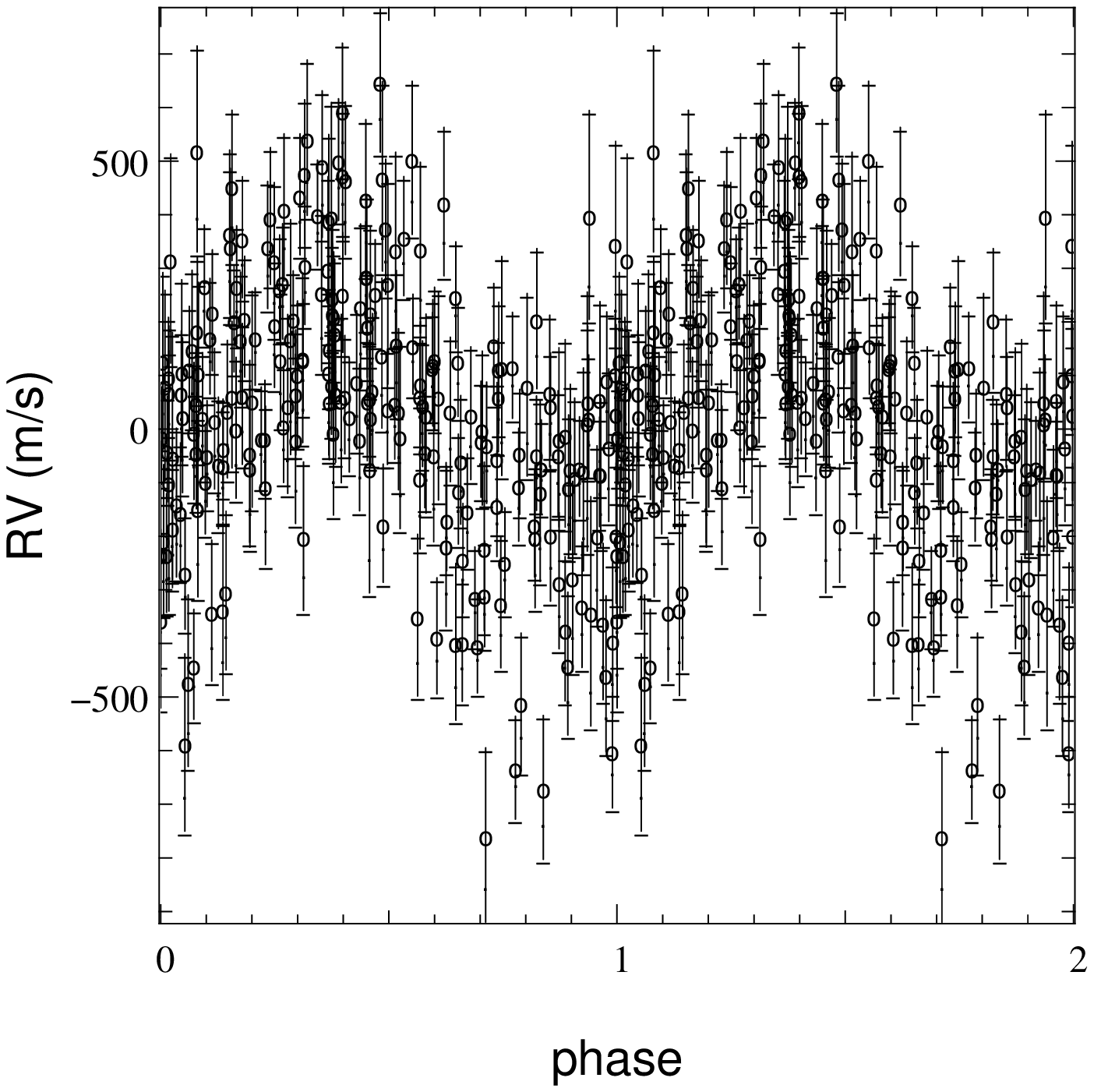}
    \includegraphics[width=0.8\columnwidth]{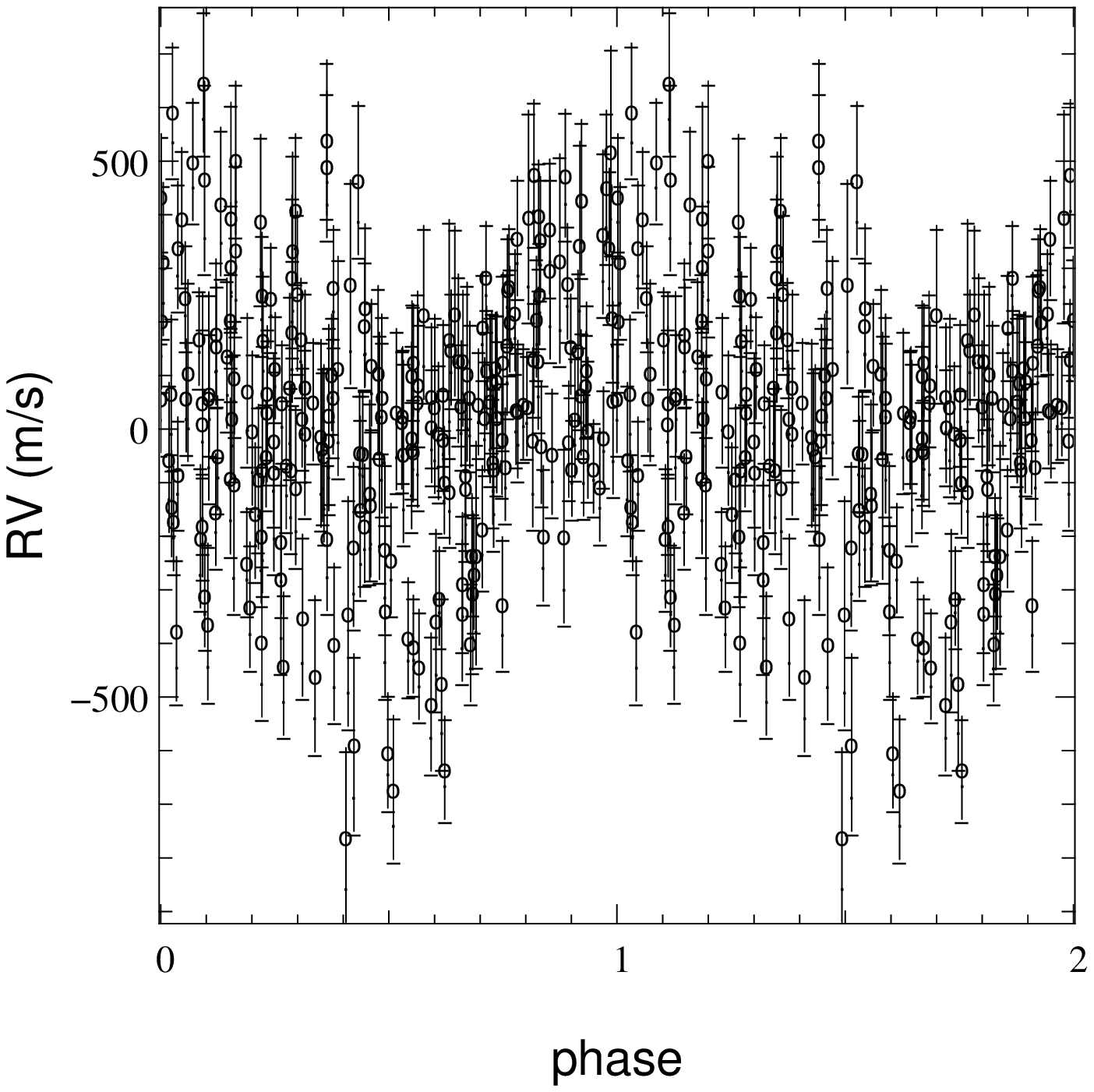}
    \caption{High frequency periodograms of the radial velocities
    obtained on $\beta$~Pictoris using the {\small HARPS} spectrograph
    (top) and the phasing of the radial velocities to the
    corresponding periods (bottom). The periodograms were obtained
    using the CLEAN algorithm (see text).  }
    \label{Har_dvft-perio_HF}
  \end{figure}

  \section{Origin of the radial velocity variations}

  \subsection{Ruling out cometary bodies}

  An origin of the radial velocity variations connected to the
  presence of the complex circumstellar disk of dust and gas has to be
  addressed, in particular a connection with the evaporating cometary
  bodies that have been proposed to explain the strong variations
  observed for some spectral lines of ionized elements such as
  \ion{Ca}{ii}, \ion{Fe}{ii}, \ion{Mg}{ii}, \ion{Al}{iii}
  (\cite{LBA00} 2000).
  
  We first computed the cross-correlation function of each
  spectrum with a binary mask taking into account only lines that correspond
  to neutral elements. In this way, we obtained a mean line for
  these neutral elements, with a better S/N than for individual
  lines. These cross-correlation functions do not show analogous
  variations to the ionized elements \ion{Ca}{ii}, \ion{Fe}{ii},
  \ion{Mg}{ii}, \ion{Al}{iii}.  Hence cometary infall does not produce
  detectable features (at the level of the cross-correlation
  functions) in the circumstellar lines of neutral elements.

  Moreover, we again computed the radial velocities taking only these
  lines of neutral elements into account. The radial velocities
  obtained were the same as previously, given uncertainties; the
  distribution of the differences between them has a dispersion of 78
  \,m\,s$^{\rm -1}$, close to uncertainties. We can then conclude that
  the radial velocity variations are unlikely to be related to the
  evaporating cometary bodies.

  \subsection{Ruling out activity}
  
  For active stars, spots on the stellar surface induce radial
  velocity variations (``jitter''), with a period equal to the star
  rotation period. The $\beta$~Pictoris rotation period is about 16
  hours. The periodogram of the radial velocities obtained with
  {\small HARPS} does not show any peak in this range of frequencies,
  see Fig.\,\ref{Har_dvft-perio} (bottom, right).  Moreover,
  $\beta$~Pictoris does not show surface abundance anomalies
  in contrast to Ap stars, which also show spectroscopic peculiarities
  attributed to magnetic activity (\cite{holweger97}).  Furthermore,
  these stars are usually slower rotators
  ($v\sin{i}$~$\leq$~120~km\,s$^{\rm -1}$, with a bulk at
  80~km\,s$^{\rm -1}$), whereas $\beta$~Pictoris $v\sin{i}$ is larger
  than 120~km\,s$^{\rm -1}$ (\cite{abt00}).  Activity should thus not
  be responsible for the variations in the radial velocities.

  \subsection{Pulsations}

  Even if our temporal sampling does not allow a detailed analysis of
  short period variations, the large number of spectra obtained with
  {\small HARPS} allows us to enhance two frequencies characteristic
  of pulsations: at 47.44$\pm0.01$~cycle\,d$^{\rm -1}$ (period of
  30.4~min) and 39.05$\pm0.01$~cycle\,d$^{\rm -1}$ (period of
  36.9~min) (Fig.\,\ref{Har_dvft-perio_HF}, top). The square root of
  the value of the pics in the periodogram stands for the radial
  velocity semi-amplitude of the corresponding radial velocity
  periodic variations: $\sqrt{0.046}$ and $\sqrt{0.029}$\,km\,s$^{\rm
  -1}$, i.e. 215 and 170\,m\,s$^{\rm -1}$, respectively, for the pics at
  47.44 and 39.05~cycle\,d$^{\rm -1}$.  The phasing of the radial
  velocities to the derived periods confirms their reality
  (Fig.\,\ref{Har_dvft-perio_HF}, bottom), as well as the
  corresponding radial velocity amplitudes (typically 200\,m\,s$^{\rm
  -1}$). The correction of an adjustment of the radial velocities
  with the superposition of two sinusoides with periods fixed to the
  above values leads to a decrease in the radial velocity dispersion
  from 252 to 182\,m\,s$^{\rm -1}$, which is still well above the
  uncertainties, 65\,m\,s$^{\rm -1}$ on average.  However, note that
  these uncertainty values suppose that the spectra are identical and
  only shifted from one to the other due to the Doppler-Fizeau
  Effect. They should be larger considering variations in the shape of
  the lines.  This adjustment is reached for values of the amplitude
  of 216 and 149\,m\,s$^{\rm -1}$, respectively, for the periods
  corresponding to 47.44 and 39.05~cycle\,d$^{\rm -1}$.

  These results agree with those obtained by \cite{Koen03}
  from dedicated photometry and
  spectroscopy. These authors indeed report detecting of at
  least 18 pulsation modes in $\beta$~Pictoris, with a large number of
  spectra spread over 2 weeks, and detecting 2 low amplitude
  ($\leq$ 1.5 mmag) pulsation modes in photometry, with frequencies
  equal to  47.44 cycle\,d$^{\rm -1}$ and 39.05 cycle\,d$^{\rm -1}$,
  namely the same as we develop here.
  We are not able to detect other high frequencies, maybe because
  our temporal sampling is not really adapted to seeking high
  frequency variations.

  The presence of pulsations in the case of $\beta$~Pictoris is not
  really surprising, as this star belongs to the left side of the
  range of B-V, where the Instability Strip intersects with the Main
  Sequence (\cite{Eyer97}). As the frequencies of the variations are
  larger than 0.25 cycle\,d$^{\rm -1}$ (periods inferior to 6.5~h)
  and the stellar mass is larger than 1.9~$M_\odot$,
  $\beta$~Pictoris probably belongs to the pulsating $\delta$~Scuti
  stars, which undergo non radial pulsations of p-mode excited by the
  $\kappa$ mechanism with \ion{He}{ii} (\cite{Handler02},
  \cite{Breger00}).

  \begin{figure}[t!]
    \hglue
    0.09\columnwidth\includegraphics[width=0.9\columnwidth]{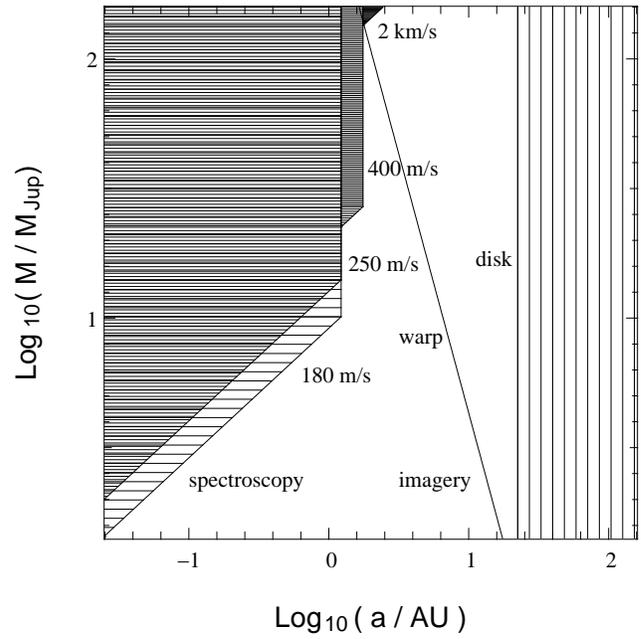}
    \caption{Domain of [mass (M) / separation (a)] where the presence
      of a planet around $\beta$~Pictoris is excluded by this
      radial velocity study (hashed zones, left): limit of 180 and
      250~m\,s$^{\rm -1}$ correspond to {\small HARPS}, 400~m\,s$^{\rm
      -1}$ to {\small CORALIE}.  The straight line indicates the
      possible characteristics of the planet responsible for the
      warped disk (\cite{mouillet97}). The disk location is also
      indicated (right).  }
    \label{Diag-exclus}
  \end{figure}

  \section{Exclusion domain for a planet around $\beta$~Pictoris}

  The periodogram of the {\small HARPS} radial velocities, now
  corrected from the variations induced by the pulsations found above,
  does not show any clear peak in a period range of 1-180
  days. Assuming a circular orbit again, we can still exclude the
  presence of a planet with a period lying typically between 1 and 350
  days (hence a separation between 0.03 and 1.2 AU), but this time
  with an induced radial velocity semi-amplitude decreasing
  to~$\approx$~180\,m\,s$^{\rm-1}$ (Fig.\,\ref{Diag-exclus}),
  which corresponds to the radial velocity dispersion after correction of
  the pulsations.

  In Fig.\,\ref{Diag-exclus}, we reproduce the planet (mass,
  separation) domain constrained by the presence of the warp
  (\cite{augereau01}, \cite{mouillet97}), the present radial velocity
  measurements (limit of 180 and 250~m\,s$^{\rm -1}$ correspond to
  {\small HARPS}, 400~m\,s$^{\rm -1}$ to {\small CORALIE}), as well as
  older ones obtained by \cite{lagrange92}; dispersion of
  2~km\,s$^{\rm-1}$, corresponding then to the achieved
  precision). The present analysis clearly constrains a new and
  important part of the domain.
  
  \section{Conclusions}

  In the frame of the search for extrasolar planets and brown dwarfs
  around early-type stars, we obtained a large number of spectra of
  $\beta$~Pictoris over several months with the {\small CORALIE} and
  {\small HARPS} spectrographs.  The radial velocities obtained
  exclude the presence of an inner giant planet in the
  $\beta$~Pictoris system. Yet, these radial velocities are
  significantly variable, and we attribute at least a part of these
  variations to pulsations of $\delta$~Scuti type.  Because of the
  effects of pulsations on the radial velocities, the stars belonging
  to the intersection of the Instability Strip and the Main Sequence,
  such as $\beta$~Pictoris, have to be carefully studied when looking for
  planets.

  \begin{acknowledgements}
    We acknowledge H.~Beust and A.~Vidal-Madjar for their fruitful
    discussions, and the Swiss National Science Foundation for its
    support of the {\small CORALIE} programmes.  We are grateful to
    ESO for the time allocation, and to the technical staff operating
    the 3.6-m telescope and the {\small HARPS} spectrograph at La
    Silla Observatory.  We acknowledge support from the French CNRS
    and the Programme National de Plan\'etologie (PNP, INSU).
  \end{acknowledgements}
  


\begin{thebibliography}{}
    
  \bibitem[Abt 2000]{abt00} Abt, H., 2000, ApJ, 544, 933
    %
  \bibitem[Augereau et al. 2001]{augereau01} Augereau, J.-C., Nelson,
    R.-P., Lagrange, A.-M., et al. 2001, A\&A, 370, 447
    %
  \bibitem[Barrado Y Navascues et al. 1999]{Barrado99} Barrado Y
    Navascues, D., Stauffer, J.-R., Song, I., et al. 1999, ApJ, 520, 123
    %
  \bibitem[Beust et al. 1990]{beust90} Beust, H., Vidal-Madjar, A.,
    Ferlet, R., et al. 1990, A\&A, 236, 202	
    %
  \bibitem[Beust \& Morbidelli 1996]{beust96} Beust, H., Morbidelli, A.,
    1996, Icar, 120, 358	
    %
  \bibitem[Beust \& Morbidelli 2000]{beust00} Beust, H., Morbidelli, A.,
    2000, Icar, 143, 170	
    %
  \bibitem[Breger et al. 2000]{Breger00} Breger, M., Montgomery, M.-H., 2000,
    ASP Conf. Ser., Vol. 210, Delta Scuti an Related
    stars. Astron. Soc. Pac., San Francisco
    %
  \bibitem[Chelli 2000]{Chelli00} Chelli, A., 2000, A\&A, 358, L59
    %
  \bibitem[Crifo et al. 1997]{crifo97} Crifo, F., Vidal-Madjar, A.,
    Lallement, R., et al. 1997, A\&A, 320, L29
    %
  \bibitem[Eyer et al. 1997]{Eyer97} Eyer, L., Grenon, M.,  1997,
    Hipp.Conf, 467E	
    %
  \bibitem[Ferlet et al. 1987]{Ferlet87} Ferlet, R., Vidal-Madjar, A.,
    Hobbs, L.-M., 1987, A\&A, 185, 267	
    %
  \bibitem[Galland et al. 2005a]{Galland05a} Galland, F., Lagrange,
    A.M., Udry, S., et al. 2005a, A\&A, in press
    %
  \bibitem[Galland et al. 2005b]{Galland05b} Galland, F., Lagrange,
    A.M., Udry, S., et al. 2005b, A\&A, in press
    %
  \bibitem[Handler et al. 2002]{Handler02} Handler, G., Balona, L.-A.,
    Shobbrook, R.-R., et al. 2002, MNRAS, 333, 262
   %
  \bibitem[Hipp]{Hipp} ESA 1997, The Hipparcos and Tycho Cat, ESA SP-1200
   %
  \bibitem[Holweger et al. 1997]{holweger97}  Holweger, H., Hempel, M.,
    van Thiel, T., et al. 1997, A\&A, 320, L49
    %
  \bibitem[Koen et al. (2003)]{Koen03} Koen, C., Balona, L.-A., Khadaroo,
    K., et al, 2003, MNRAS, 344, 1250
    %
  \bibitem[Lagrange et al. 1988]{lagrange88} Lagrange-Henri, A.-M.,
    Vidal-Madjar, A., Ferlet, R., 1988, A\&A, 190, 275	
    %
  \bibitem[Lagrange et al. (1992]{lagrange92} Lagrange-Henri, A.-M.,
    Gosset, E., Beust, H., et al. 1992, A\&A, 264, 637
    %
  \bibitem[Lagrange et al.]{LBA00} Lagrange, A.-M., Backman, D.-E., Artymowicz,
    P. , 2000,
    in Protostars and Planets IV (Book - Tucson: University of Arizona Press;
    eds Mannings, V., Boss, A.P., Russell, S. S.), p. 639
    %
  \bibitem[Lagrange et al. 2004]{LA04} Lagrange, A.-M., Augereau, J.-C.,
    2004, in Planetary systems and planets in systems, ISSI workshop
    %
  \bibitem[Lamers et al. 1997]{lamers97}  Lamers, H.,
    Lecavelier Des Etangs, A., Vidal-Madjar, A., 1997, A\&A, 328, 321
    %
  \bibitem[Lecavelier Des Etangs et al. 1995]{lecavelier95}  Lecavelier Des
    Etangs, A., Deleuil, M., Vidal-Madjar, A., et al. 1995, A\&A, 299, 557
    %
  \bibitem[Lecavelier Des Etangs et al. 1996]{lecavelier96}  Lecavelier Des
    Etangs, A., Vidal-Madjar, A., Ferlet, R., 1996, A\&A, 307, 542	
    %
  \bibitem[Lecavelier Des Etangs et al. (1997)]{lecavelier97}  Lecavelier Des
    Etangs, A., Vidal-Madjar, A., Burki, G., et al. 1997, A\&A, 328, 311
    %
  \bibitem[Mathias et al. 2004]{Mathias04} Mathias, P., Le Contel, J.-M.,
  Chapellier, E., 2004, A\&A, 417, 189
    %
  \bibitem[Mouillet et al. 1997]{mouillet97} Mouillet, D., Larwood,
    J.-D., Papaloizou, J.-C.-B., et al. 1997, MRAS, 292,  896
    %
  \bibitem[Pepe et al. 2002]{Pepe02} Pepe, F., Mayor, M., Rupprecht, G., et
    al. 2002, The ESO Messenger, 110, 9 
    %
  \bibitem[Roberts et al. 1987]{Roberts87} Roberts, D.-H., Lehar, J.,
  Dreher, J. W., 1987, AJ, 93, 968
  %
  \bibitem[Smith and Terrile 1984]{ST84} Smith, B.-A., Terrile, R.-J.,
    1984, Sci, 226, 1421
    %
  \bibitem[Vidal-Majar et al. (1998)]{Majar98} Vidal-Madjar, A.,
    Lecavelier des Etangs, A., Ferlet, R., 1998, P\&SS, 46, 629	
    %
 
  \end{thebibliography}
\end{document}